\documentclass[aps,eqsecnum,preprint,floats,nofootinbib]{revtex4}
\usepackage{graphicx}
\usepackage{epsfig}

\begin{document}
\def\be{\begin{eqnarray}}
\def\en{\end{eqnarray}}
\def\ph{\phantom}
\def\non{\nonumber}
\def\jpsi{J/\psi}
\newcommand{\bo}{B^0}
\newcommand{\bob}{\bar{B^0}}
\newcommand{\bq}{B_q}
\newcommand{\bqb}{\bar{B_q}}
\newcommand{\bd}{B_d}
\newcommand{\bs}{B_s}
\newcommand{\bsb}{\bar{B_s}}
\newcommand{\ks}{K_S}
\newcommand{\kl}{K_L}
\def\bbarb{B^0 -\bar B^0}
\def\OL{\overline}
\def\la{\langle}
\def\ra{\rangle}
\def\pp{{\prime\prime}}
\def\nc{N_c^{\rm eff}}
\def\vp{\varepsilon}
\def\hep{\hat{\varepsilon}}
\def\drho{\bar\rho}
\def\deta{\bar\eta}
\def\a{{\cal A}}
\def\B{{\cal B}}
\def\c{{\cal C}}
\def\d{{\cal D}}
\def\e{{\cal E}}
\def\p{{\cal P}}
\def\t{{\cal T}}
\def\B{{\cal B}}
\def\L{{\cal L}}
\def\P{{\cal P}}
\def\S{{\cal S}}
\def\T{{\cal T}}
\def\C{{\cal C}}
\def\A{{\cal A}}
\def\E{{\cal E}}
\def\V{{\cal V}}
\def\CP{{\it CP}~}
\def\CPP{{\it CP}}
\def\up{\uparrow}
\def\dw{\downarrow}
\def\vma{{_{V-A}}}
\def\vpa{{_{V+A}}}
\def\smp{{_{S-P}}}
\def\spp{{_{S+P}}}
\def\lrpartial{\buildrel\leftrightarrow\over\partial}
\def\J{{J/\psi}}
\def\3bar{{\bf \bar 3}}
\def\6bar{{\bf \bar 6}}
\def\10bar{{\bf \ov{10}}}
\def\ov{\overline}
\def\Lqcd{{\Lambda_{\rm QCD}}}
\def\pr{{Phys. Rev.}~}
\def\prl{{ Phys. Rev. Lett.}~}
\def\pl{{ Phys. Lett.}~}
\def\np{{ Nucl. Phys.}~}
\def\zp{{ Z. Phys.}~}
\def\hep{hep-ph}
\def\Ref{\bibitem}
\def\lsim{ {\ \lower-1.2pt\vbox{\hbox{\rlap{$<$}\lower5pt\vbox{\hbox{$\sim$}
}}}\ } }
\def\gsim{ {\ \lower-1.2pt\vbox{\hbox{\rlap{$>$}\lower5pt\vbox{\hbox{$\sim$}
}}}\ } }

\font\el=cmbx10 scaled \magstep2{\obeylines\hfill BNL-HET-06/18}

\font\el=cmbx10 scaled \magstep2{\obeylines\hfill \today}

\vskip 1.5 cm

\centerline{
  \large \bf \boldmath
  Null tests of the Standard Model at an
  International Super $B$ Factory
}  
\bigskip
\centerline{\bf Tim Gershon$^1$ and Amarjit Soni$^2$}
\medskip
\centerline{
  $^1$ Department of Physics, University of Warwick,
}
\centerline{
  Coventry, CV4 7AL, UK
}
\medskip
\medskip
\centerline{
  $^2$ Physics Department, Brookhaven National Laboratory
} 
\centerline{
  Upton, New York 11973, USA
}
\medskip

\bigskip
\bigskip
\centerline{\bf Abstract}
\bigskip
\small

In light of the results from the $B$ factories, 
which clearly show that 
the Cabibbo-Kobayashi-Maskawa mechanism
is the dominant source of the observed
$CP$ violation in $K$ and $B$ physics, only small deviations due to 
sources of $CP$ violation beyond the Standard Model are likely. 
Therefore, in the quest for New Physics,
{\it null tests} of the Standard Model become increasingly important.  
Motivated by these considerations,
we describe a number of approximate null tests of the Standard Model.
These tests provide several theoretically clean approaches to 
searching for new physics in the $B$ system.
We find that in many cases, the requisite sensitivity can only
be achieved with an International Super $B$ Factory, 
with luminosity around $10^{36} {\rm cm}^{-2}{\rm s}^{-1}$.

\eject

\section{Introduction}
The $B$ factories, BaBar~\cite{babar} and Belle~\cite{belle}, 
have helped us attain an important milestone in Particle
Physics with respect to our understanding of $CP$ violation
phenomena: measurements of the angles ($\phi_1,\phi_2,\phi_3 / \alpha,\beta,\gamma$)
and sides of the Unitarity Triangle are in
very good agreement~\cite{ckmfit,utfit,hfag}
with the expectations from the Standard Model (SM) \textendash{} 
the CKM~\cite{ckm} paradigm. 
Therefore, effects of the Beyond Standard Model (BSM) $CP$-odd phase(s)
are likely to be a perturbation. 
In fact the SM itself teaches us an important lesson in this regard.  
Recall that we know now that the SM CKM phase is ${\cal O}\left( 1 \right)$
and yet it causes large $CP$-odd effects only in $B$ physics. 
In top physics the SM $CP$ violation effects
are completely negligible~\cite{ehs91,abes_pr}. 
In charm physics they are also expected to be very small~\cite{charm}. 
In $K$ decays the observed $CP$ asymmetries are 
$\epsilon_K \approx 10^{-3}$ and $\epsilon' \approx 10^{-5}$,
both extremely small~\cite{PDB}. 
So, while compelling theoretical rationale dictates the existence of 
BSM $CP$-odd phase(s), even if such a phase is ${\cal O}\left( 1 \right)$, 
its effect on $B$ physics need not be large.
In fact, given the already existing constraints, 
such effects are likely to be small. 
Indeed there is no reason to suggest that
they may not be as small as $\epsilon_K \approx 10^{-3}$. 
It is therefore very important that we sharpen our ability 
to search for small deviations.   

To facilitate such searches we need: 
\vspace{-0.4\baselineskip}
\begin{enumerate}\addtolength{\itemsep}{-0.6\baselineskip}
\item{} Clean predictions from theory.
\item{} Precise experimental measurements requiring large numbers of $B$ mesons.
\end{enumerate}
 
Null tests of the SM, {\it i.e.} asymmetries that are forbidden 
or expected to be small can be particularly useful in searching for
small deviations. 
The construction of strict null tests is in line with item 1) above. 
Since $CP$ is not a symmetry of the SM, we cannot have exact null tests 
but rather we must settle for approximate null tests (ANTs). 
One such null test that has been much discussed recently 
(and is also considered below)
involves comparing the time-dependent $CP$ asymmetry in 
penguin-dominated modes such as $B_d \to \phi K_S, \eta^\prime K_S,$ {\it etc.},
with the Standard Model reference $B_d \to J/\psi K_S$. 
A rough theoretical expectation for this ANT is that~\cite{gw97,ls97,giw97}:
\be
\left| \Delta S \right| = 
\left| S_{\rm penguin} - S_{\psi K_S} \right| \lsim 
{\cal O}\left( \lambda^2 \right) \approx 5\%.
\en
Experimentally this expectation seems to be off by about 2$\sigma$~\cite{hfag},
while recent theoretical treatments, discussed in the next section,
have reduced the theory error of this ANT for some specific modes.
While more precise results are keenly anticipated,
it is also clearly important to search for corroborative evidence.
This motivates us to develop additional null tests that 
are as strict as possible.  
As will be shown, there are several ANTs with 
much smaller theoretical uncertainties than $\Delta S$.

In fact, $B$ physics offers a plethora of ANTs.
Here we do not discuss them all in detail, 
but instead focus attention on a selection of examples.
For these, we review the phenomenology, 
and consider the experimental possibilities available.
In many cases, we find that precise tests of the SM can only be achieved
from a high luminosity $e^+e^-$ machine.
Thus a concerted world wide effort for an 
International Super $B$ Factory with luminosity 
$\gsim 10^{36}{\rm cm}^{-2}{\rm s}^{-1}$ is highly desirable.

\section{Some examples of Null tests}

\subsection{\boldmath
  Time-dependent $CP$ asymmetry in penguin-dominated modes}

As mentioned above, this test has been much in the news in the last 
few years as there seems to be an apparent $\approx 2 \sigma$
deviation from the SM. 
Prior to summer 2006,
most of the effect seemed to originate
from BaBar's measurement of the $S$ parameter 
(the time-dependent $CP$ asymmetry) in $B^0 \to \eta^\prime K^0$~\cite{babar_etaprk0},
which gave $- \eta_{CP} S_{\eta^\prime K^0} = 0.36 \pm 0.13 \pm 0.03$,
to be compared to the Belle measurement (at the same time) of
$0.62 \pm 0.12 \pm 0.04$~\cite{belle_etaprk0},
and an average of $0.50 \pm 0.09$.
[The first error is statistical and the second is systematic.
The parameter $\eta_{CP}$ gives the $CP$ eigenvalue of the final state,
which is $-1$ for $\eta^\prime K_S$ and $+1$ for $\eta^\prime K_L$.]
New results were presented at ICHEP 2006, and more recently, giving
(BaBar) $0.58 \pm 0.10 \pm 0.03$~\cite{babar_etaprk0_2006} and
(Belle) $0.64 \pm 0.10 \pm 0.04$~\cite{belle_etaprk0_2006}, 
with an average of $0.61 \pm 0.07$.
These results have to be compared with an average 
$- \eta_{CP} S_{J/\psi K^0} = 0.675 \pm 0.026$~\cite{hfag}.
The latest results considerably reduce the discrepancy in the $\eta^\prime K^0$ mode.
Nonetheless, there remains an clear trend, 
in that all measurements of 
time-dependent $CP$ asymmetries in penguin-dominated modes
yield central values below the value from $J/\psi K^0$.

When this class of null tests were first proposed~\cite{gw97,ls97,giw97}
it was estimated that $\Delta S \lsim O(\lambda^2) \approx 5\%$.
In view of the possible experimental hint of a deviation,  
several theoretical approaches have recently been used to
to re-examine this expectation. In particular,
effects of final state interactions were completely ignored
in the original estimates. In view of the experimental
observations of large direct $CP$ in $K^+\pi^-$~\cite{kpi_dcpv}
and several other modes~\cite{hfag}, 
it is clear that final state interaction (FSI) effects in exclusive $B$ decays 
can be sizeable~\cite{ccs1}. 
Therefore, their influence on the $S$ parameters 
of the penguin-dominated modes of
interest was systematically investigated~\cite{ccs2,ccs3}. 
It seems that $\eta^\prime K^0$, $\phi K^0$ and $K_S K_S K^0$~\cite{gh}
are the cleanest channels in this approach. 
Table~\ref{penguin} compares the results for $\Delta S$ from 
some of the theoretical studies~\cite{ccs2,ccs3,mb05,bhnr05,wz06}.  
While three of the studies in the table use QCDF~\cite{qcdf}, 
it is interesting to note that the first application
of soft collinear effective theory (SCET)~\cite{scet} 
to the calculation of $\Delta S$ has been made~\cite{wz06}, 
though it is only, for now, for $\eta^\prime  K^0$.

In addition to the calculations discussed above, 
bounds on the possible deviation in the Standard Model
have been obtained using flavour SU(3) symmetry~\cite{glnq2003,grz2004}.
These bounds correlate the values of $\Delta S$ with the 
size of the direct $CP$ violation parameters in each channel.
Since these bounds rely on measurements of channels involving the 
flavour SU(3) partners of the final states of interest,
they tend to be less constraining than the explicit calculations.
However, they can be improved with additional data.

\begin{table}[htbp]
  \caption{Some expectations for $\Delta S$ in the cleanest modes.} 
  \label{penguin}
  \begin{center}
    \begin{tabular}{|@{\hspace{3mm}}c@{\hspace{3mm}}|@{\hspace{3mm}}c@{\hspace{3mm}}|@{\hspace{3mm}}c@{\hspace{3mm}}|@{\hspace{3mm}}c@{\hspace{3mm}}|@{\hspace{3mm}}c@{\hspace{3mm}}|}
      \hline
      Mode & QCDF+FSI~\cite{ccs2,ccs3} & QCDF~\cite{mb05} & QCDF~\cite{bhnr05} 
      & SCET~\cite{wz06} \\ 
      \hline
      $\eta^\prime K^0$ & $0.00^{+0.00}_{-0.04}$ & $0.01 \pm 0.01$ & $0.01 \pm 0.02$ & $-0.019 \pm 0.009$ \\
      & & & & $-0.010 \pm 0.001$ \\
      \hline
      $\phi K^0$  & $0.03^{+0.01}_{-0.04}$ & $0.02 \pm 0.01$ & $0.02 \pm 0.01$ &  \\
      \hline
      $K_S K_S K^0$ & $0.02^{+0.00}_{-0.04}$ & & & \\
      \hline
    \end{tabular}
  \end{center}
\end{table}

In passing, we also want to mention~\cite{as_lathuile,ccs2}
that although the sign of the central value of $\Delta S$
found in several theoretical models for many of the modes of interest
(see Table~\ref{penguin}) 
tends to be opposite to that found experimentally, 
the theory errors are sufficiently large that this
by itself is not a reliable sign of NP. 
Besides, in the recent SCET based calculation~\cite{wz06}
it was found that for the $\eta^\prime K^0$ case the sign of $\Delta S$ is different 
from the three model calculations in Table~\ref{penguin}
and in fact is the same as seen experimentally~\cite{zl_thanks}.

In order to consider how much data is necessary to probe this null test,
we need to assess the theoretical uncertainty in the predictions.
In each of the modes listed in Table~\ref{penguin}
the typical uncertainty is about $0.02$.
However, as mentioned above, 
there is some spread between predictions coming from different models, 
and there are also uncertainties in the calculations themselves, 
so that this value may be too aggressive at the present time.  
Hopefully, improved understanding of the $B$ decay dynamics will 
lead to more precise predictions in future.
However, it is clear that values of $|\Delta S| >0.10$ are extremely difficult
to reconcile with the SM CKM paradigm,
underpinning the original predictions~\cite{gw97,ls97,giw97}.
Note, however, that this is not necessarily true in some other 
hadronic decay channels being used to study the $b \to s$ penguin
transition (such as $\rho^0K^0$, $\omega K^0$ and $\eta K^0$)~\cite{ccs2,ccs3,wz06}.

The above discussion notwithstanding, 
let us consider how much data will be necessary to reach a precision of $0.02$.
In order to do so, we can simply extrapolate from the existing measurements.
The average values have uncertainties of 
$0.07$ ($\eta^\prime K^0$), $0.18$ ($\phi K^0$) and $0.21$ ($K_SK_SK^0$)~\cite{hfag}.
These come from about 900 million $B\bar{B}$ pairs
(535 million from Belle and 347 million (384 million for $\eta^\prime K^0$) from BaBar).
It is worthwhile to note that the BaBar result on $\phi K^0$ is extracted
from a time-dependent Dalitz plot analysis 
of $B^0 \to K^+K^-K^0$~\cite{babar_phiK0_2006},
which provides additional sensitivity.
To reach the level of the theory uncertainty,
one would need over $10^{10}$ $B\bar{B}$ pairs for $\eta^\prime K^0$,
and about $5 \times 10^{10}$ $B\bar{B}$ pairs for $\phi K^0$ and $K_SK_SK^0$.
One may consider the potential of a hadronic machine to address these modes.
At present, it appears that $\phi K_S$ is difficult, but not impossible
to trigger and reconstruct, due to the small opening angle in
$\phi \to K^+K^-$ in the hadronic environment;
$\eta^\prime K_S$ is challenging since neutral particles are involved in the 
$\eta^\prime$ decay chain; for $K_SK_SK_S$ meanwhile, 
there are no charged tracks originating from the $B$ vertex,
and so both triggering and reconstruction seem highly complicated.
Modes containing $K_L$ mesons in the final state 
may be considered impossible to study at a hadron machine.
Thus, these modes point to a Super $B$ Factory,
with integrated luminosity of at least $50 \ {\rm ab}^{-1}$.

\subsection{
  A class of inclusive hadronic $B$ decays as null tests of the SM
}

Ref.~\cite{snz05} shows that the SM CKM paradigm predicts completely
negligible partial width differences (PWDs) 
in
$B^{\pm} \to M^0 (\bar{M}^0) X^{\pm}_{s+d}$
where $M^0$ has energy close to $m_b/2$ 
({\it i.e.} is in the end-point region~\cite{cklz06}), 
and is either
1) an eigenstate of $s \leftrightarrow d$ switching symmetry,
{\it e.g.} $K_S$, $K_L$, $\eta^\prime$ or any charmonium state,
or 2) $M^0$ and $\bar M^0$ are related by the $s \leftrightarrow d$ transformation,
{\it e.g.} $K^0$, $\bar{K}^{*0}$, $D^0$.
In these precision tests the PWD
suffers from double suppression,
{\it i.e.} ${\cal O}(\lambda^2)$ suppression comes from CKM-unitarity,  
and U-spin symmetry of QCD causes an additional suppression
which is naively expected to be ${\cal O}(m_s/\Lambda_{\rm QCD})$. 
Briefly this can be understood as follows. 
Recall that CKM unitarity allows us to write the $\Delta S=1$ decay width as
\be
\Gamma(B^-\to M^0 X_s^-) = | \lambda_c^{(s)} A_c^s + \lambda_u^{(s)} A_u^s |^2,
\en
where $A_{u,c}^s$ denote the terms in the amplitude proportional to
corresponding CKM matrix elements
$\lambda_c^{(s)}=V_{cb} V_{cs}^*\sim \lambda^2$ and
$\lambda_u^{(s)}=V_{ub} V_{us}^*\sim \lambda^4$
(with $\lambda = \sin \theta_c = 0.22$), 
the corresponding $\Delta S=1$ PWD is
\be
\Delta \Gamma^s 
& = & \Gamma(B^- \to M^0 X_s^-) - \Gamma(B^+ \to M^0 X_s^+) \\
& = & - 4 J \Im[A_c^{s} A_u^{s*}],
\en  
with
$J = \Im[\lambda_c^{(s)}\lambda_u^{(s)*}] = -\Im[\lambda_c^{(d)}\lambda_u^{(d)*}]$, 
the Jarlskog invariant. 
Note that $A_{u,c}^s$ are complex due to strong phases.
Similarly for the $\lambda^2$ suppressed $\Delta S=0$
decay
\be
\Delta \Gamma^d
& = & \Gamma(B^- \to M^0 X_d^-) - \Gamma(B^+ \to M^0 X_d^+) \\
& = & 4 J \Im[A_c^{d} A_u^{d*}].
\en  
The transformation $s \leftrightarrow d$ exchanges $X_s$ and $X_d$ final states, 
while it has no effect on $B^\pm$ and $M^0$ states.
In the limit of exact U-spin thus $A_{u,c}^s = A_{u,c}^d$, 
giving a vanishing PWD in 
flavour untagged inclusive decay~\cite{snz05,js,mg,hm}
\be
\label{eq:dg_splusd}
\Delta \Gamma^{s+d} = \Delta \Gamma^s + \Delta \Gamma^d & = - 4 J \Im[A_c^{s} A_u^{s*} - A_c^{d} A_u^{d*}] = 0.
\en
To the extent that U-spin is a valid symmetry of strong interactions the
observable $\Delta \Gamma^{s+d}$
constitutes a null test of SM. Quite generally the breaking can be
parameterised
\be
\Delta \Gamma^{s+d} \equiv \delta_{s \leftrightarrow d} \Delta \Gamma^s,
\en
leading to an expectation for the $CP$ asymmetry of the decay into
untagged light flavour~\cite{snz05} 
\be
\label{eq:acp_splusd}
{\cal A}_{CP}^{s+d} = 
\frac{\Delta \Gamma^s + \Delta \Gamma^d}{\bar{\Gamma}^{s+d} + {\Gamma}^{s+d}} \sim \delta_{s \leftrightarrow d} \lambda^2,
\en
where in the last relation use of the facts that
$\Delta\Gamma^d \sim \Delta\Gamma^s \sim \lambda^2 \Gamma^s$ and $\Delta\Gamma^d \sim \Gamma^d$ have been made. 
The size of the U-spin breaking parameter $\delta_{s \leftrightarrow d}$ 
is channel dependent with an order of magnitude expectation 
$\delta_{s \leftrightarrow d} \sim m_s/\Lambda_{\rm QCD} \sim 0.3$. 
Similarly, the size of the $\Delta S=1$ $CP$ asymmetry
$\Delta\Gamma^s/(\Gamma^s + \bar{\Gamma}^s) \simeq 2 (J/\lambda_c^{(s)2}) \Im[A_u^{(s)}/A_c^{(s)}]$ 
depends on the decay channel through the ratio of the two amplitudes. 
The sizes of the asymmetries, ${\cal A}_{CP}^{s+d}$,
for choices of $M^0$ such as $D^0 + \bar{D}^0$, $\eta^\prime$, $K^0 + \bar{K}^0$,
are found to be well below $1\%$~\cite{snz05,cklz06}.

Note that with respect to the direct $CP$ asymmetry in
final states containing $\eta^\prime$ mesons, this approach~\cite{snz05} 
represents a considerable theoretical improvement
over the semi-inclusive test with $X_s$ final state that existed in the 
literature~\cite{eta_xs}.  

Since these tests apply to inclusive final states,
certain experimental considerations are necessary.
Although the branching fractions to inclusive final states can be large,
it may be difficult to perform the reconstruction with high efficiency.
Broadly, two strategies are available: 
1) semi-inclusive, and 2) fully inclusive.

In semi-inclusive analysis, 
(sometimes also referred to as ``pseudo-reconstruction'')
the inclusive state is approximated by a sum of exclusive modes.
Generally, only the modes with relatively low multiplicities,
and low numbers of neutral final state particles, are reconstructable.
Therefore, only a fraction of the inclusive state can be included,
and knowledge of that fraction requires theoretical input
(on the fragmentation of the $X_{s,d}$ system).
For the semi-inclusive reconstruction of $X_{s,d}$,
we have to take into account that the reconstructed fraction of 
the $X_s$ system ($f_s$) will differ from that for the $X_d$ ($f_d$).
This results in the following modification to Eq.~\ref{eq:dg_splusd}:
\be
\left. \Delta \Gamma^{s+d} \right|_{\rm meas} = 
f_s \Delta \Gamma^s + f_d \Delta \Gamma^d = (f_s - f_d) \Delta \Gamma^s + f_d \Delta \Gamma^{s+d},
\en
and hence a nonzero PWD can be induced even with exact U-spin symmetry.
In case $f_s$ and $f_d$ are much below unity,
the associated uncertainty may be hard to quantify,
since these quantities depend on the fragmentation of the $X_{s,d}$ system.
Similar modifications are necessary to the direct $CP$ asymmetry 
of Eq.~\ref{eq:acp_splusd}.
This is a limitation of semi-inclusive analyses.

The semi-inclusive strategy does hold the advantage that it may 
be attempted at any facility producing $B$ mesons.
Furthermore, backgrounds may not be too large,
at least for some of the exclusive modes being summed,
and background from neutral $B$ decays should be small.
However, as the multiplicity increases, so does the background,
and it may be the case, in a hadronic environment,
that a purely exclusive approach is preferable \textendash{} 
the larger associated theoretical uncertainty notwithstanding.

The semi-inclusive approach has successfully been used 
in a number of analyses to date~\cite{btosll:babar,btosll:belle,semi-inc}.
By contrast, the fully inclusive analysis suffers from much larger
backgrounds, including those from other $B$ decays.
The strategy in this case would be to require a high momentum $M^0$,
and then make vetoes and cuts to attempt to reduce the background.
Due to difficulties controlling the background,
this technique has in the past had limited (though notable) 
success~\cite{full-inc,babar:acp-sdgamma}.

The possibility of the Super $B$ Factory opens the door for an alternative
approach: that of fully inclusive analysis on the recoil.
This technique takes advantage of the $e^+e^- \to \Upsilon(4S) \to B^+B^-$ 
production chain.
One charged $B$ is reconstructed, typically in a 
hadronic decay mode such as $B^- \to D^0 \pi^-$, 
and then $M^0$ is searched for in the remainder of the event.
One can then reconstruct the four-momentum of the $X_{s,d}$ system,
and make additional selections if necessary, for example to remove 
decays containing open charm.
This approach reduces the background to an essentially negligible level,
but carries a high price in terms of efficiency.
At present, the $B$ factories achieve efficiencies of ${\cal O}(10^{-3})$ 
to reconstruct one $B$ meson in a hadronic decay mode,
and slightly higher if semileptonic decays are also included.
Taking as an example $M^0 = \eta^\prime$, 
assuming a $10\%$ efficiency for $\eta^\prime$ reconstruction
and taking ${\cal B}(B^+ \to \eta^\prime X_{s,d}^+) \sim 5 \times 10^{-4}$~\cite{bonvicini03},
we find that in order to reach a sensitivity to ${\cal A}_{CP}^{s+d}$ of $1\%$,
more than $10^{12}$ $B\bar{B}$ pairs are necessary.
Therefore, a Super $B$ Factory with luminosity in excess of 
$10^{36}{\rm cm}^{-2}{\rm s}^{-1}$ would be needed.

\subsection{\boldmath
  Direct $CP$ asymmetry in inclusive radiative $B$ decays}

Let us now briefly mention some other inclusive decays that provide
very important null tests of the SM: namely ${\cal A}_{CP}(B \to X_{s(d)}\gamma)$.
These have been theoretically studied and emphasized for a very long
time and their predictions in the SM are rather well-established,
namely around $0.5\%$ for $X_s$ 
and around $-10\%$ for $X_d$~\cite{js,kn,ksw,hlp03}.
Experimentally, BaBar~\cite{babar:acp-sgamma} have measured 
${\cal A}_{CP}(B \to X_s\gamma) = 0.025 \pm 0.050 \pm 0.015$
using $89 \times 10^6$ $B\bar{B}$ pairs,
while Belle~\cite{belle:acp-sgamma} have measured 
${\cal A}_{CP}(B \to X_s\gamma) = 0.002 \pm 0.050 \pm 0.030$
using $152 \times 10^6$ $B\bar{B}$ pairs.
Both analyses reconstruct $X_s$ from a sum of exclusive states.
The first errors are statistical and the second systematic.
Taking the theoretical uncertainty to be $0.2\%$~\cite{hlp03},
we see that approximately 350 times more data, 
{\it i.e.} around $10^{11}$ $B\bar{B}$ pairs,
will be necessary to drive the experimental errors down to this level.
Note that this simple extrapolation assumes that 
systematic errors can also be reduced,
and neglects uncertanties due to the hadronization of the $X_s$ system.

The branching fraction of $B \to X_d \gamma$ is lower by about a factor of 20,
compared to $B \to X_s \gamma$,
however the expected asymmetry is much bigger (by the same factor).
Therefore, in principle the measurement of this asymmetry may
become accessible with a comparable amount of data to the $X_s$ case.
However, in this case, the $B \to X_s \gamma$ decays provide a background
to the $B \to X_d \gamma$ signal, which complicates the analysis.
Since such measurements have not yet been attempted, 
it is not possible to quantify the effect on the sensitivity.
However, it is reasonable to assume that studies of $b \to d\gamma$ transitions
will be statistics limited even with Super $B$ Factory luminosities.

Following from this, let us in passing mention that the SM makes
an even cleaner prediction that the sum of the partial width differences
in $B \to X_s \gamma$ and $B \to X_d \gamma$ should vanish~\cite{js,hlp03}.  
This prediction is exact in the U-spin limit ($m_s = m_d$),
and corrections to it are expected to be very small~\cite{hlp03}.
In particular, in the end-point region, the flavour-breaking is 
suppressed and is $O(m_s \Lambda_{\rm QCD}/m_b^2)$~\cite{ckl2005}. 
This therefore provides a very interesting null result.
Since both $X_s \gamma$ and $X_d \gamma$ decays can now be treated as signal,
the above experimental problems no longer arise,
therefore this measurement is considerably simpler.
A first measurement of this inclusive direct $CP$ asymmetry has been
made by BaBar~\cite{babar:acp-sdgamma}, who measure
${\cal A}_{CP}(B \to X_{s+d}\gamma) = -0.11 \pm 0.12 \pm 0.02$
based on $89 \times 10^6$ $B\bar{B}$ pairs and using a fully inclusive
analysis, with a requirement that the decay of the other $B$ in the event
contains a high momentum lepton to reduce background.
Taking the theoretical uncertainty to be 
$0.2\%$ in this case, we see that if the uncertainty scales with 
luminosity, it will take over $10^{11}$ $B\bar{B}$ pairs to reach
the level of the theory error.
It will be interesting to see if refinements in the analysis techniques
can lead to faster reductions in the experimental uncertainty.
We note that, as before, inclusive measurements can not be performed 
in a hadronic environment, and therefore require a Super $B$ Factory.

\subsection{\boldmath 
  Direct $CP$ asymmetries in $B \to X_{s(d)} l^+ l^-$}

Since we have discussed radiative $B$ decays,
it is worthwhile to also mention the case where the photon 
is replaced with a dilepton pair.
Again, this reaction has been intensively studied for a very
long time and, as in the case of real photons, 
the branching fractions and the direct $CP$ asymmetries 
are of great interest for testing the SM and for searching
for New Physics~\cite{ah1999}. 
In addition, in this case the forward-backward asymmetry
is a very clean and sensitive observable~\cite{lmss99,gh03}.
The forward-backward asymmetry is defined by
$A_{\rm FB} \equiv \frac{N_{\rm F} - N_{\rm B}}{N_{\rm F} + N_{\rm B}}$,
where $N_{\rm F(B)}$ is the number of decays with the $B$ meson moving in
the forward (backward) hemisphere with respect to the direction of
motion of the positively charged lepton in the rest frame 
of the dilepton pair. 
The unique feature of $A_{\rm FB}$ is that in the SM,
as a function of the dilepton invariant mass ($s_0$), it vanishes
around $s_0 = 0.14 \ {\rm GeV}^2$ whereas in extensions of the SM
the location of this zero can be quite different and in some models
may not even exist. 
Since this $s_0$ represents a small region
of the total available kinematic range ($\approx 20 \ {\rm GeV}^2$)
and the total branching ratio in the SM is around $5 \, \times \, 10^{-6}$, 
detailed measurments of $A_{\rm FB}$ will require 
$\gsim 10^{10}$ $B$ pairs, {\it i.e.} a Super $B$ Factory~\cite{superb}.  

To date, studies of the inclusive $b \to s l^+l^-$ transition have yielded
only branching fraction measurements~\cite{btosll:babar,btosll:belle},
and a first measurement of the direct $CP$ asymmetry,
${\cal A}_{CP} = -0.22 \pm 0.26 \pm 0.02$ from BaBar,
based on $89 \times 10^6$ $B\bar{B}$ pairs and 
using a semi-inclusive analysis~\cite{btosll:babar}.
This can be compared to the SM prediction of $(-0.2 \pm 0.2)\%$,
and therefore this measurement will be statistics limited even with 
$10^{12}$ $B\bar{B}$ pairs, unless improvements to the analysis
can be achieved.
No measurements exist of $b \to d l^+l^-$, but as for $b \to d\gamma$,
we may expect that useful measurements can be carried out 
with a similar amount of data as the $b \to s$ case.
Since the PWDs for $X_s l^+l^-$ and $X_d l^+l^-$ cancel,
as before the combined $CP$ asymmetry should be an extremely precise 
null test, which only a Super $B$ Factory can test.

The exclusive counterparts, 
especially those with $K$ or $K^*$ replacing the hadronic $X_s$ system, 
are also interesting and allow somewhat easier experimental handles 
on the observables of interest:
branching ratios, direct $CP$ violation and forward-backward asymmetries.
Measurements of all of these currently exist, albeit with 
large statistical errors~\cite{babar:kll,belle:kll}.
Unfortunately the theory predictions for the exclusive modes are less clean.  
For all these radiative channels, 
we expect that certain exclusive modes can be 
very well studied at a hadronic machine, 
but that the theoretically clean inclusive analysis requires 
the environment of an $e^+e^-$ Super $B$ factory.

\subsection{The isospin sum-rule} 

Recently, several studies~\cite{bfrs04,bhlds04} 
have suggested that there may be signs of new physics in 
the electroweak penguins (EWP) sector. 
Hadronic issues such as final state interactions 
can make this completely non-trivial to decipher.
Whether or not EWP are seeing NP may be easier to discuss
once we unambiguously establish the presence of EWP in hadronic modes.
For this purpose a sum rule was proposed~\cite{ans97,mg05} 
\be
\Sigma\left(\Delta(\pi K)\right) \equiv
2 \Delta(\pi^0 K^+) - \Delta(\pi^+ K^0) - \Delta(\pi^- K^+) + 2 \Delta(\pi^0 K^0) = 0 \, ,
\en
from which a more convenient relation can easily be obtained
\be
\begin{array}{lcr}
  \multicolumn{2}{l}{
    \Sigma\left({\cal A}_{CP}(\pi K)\right) \equiv
    2 {\cal A}_{CP}(\pi^0 K^+) 
    \frac{{\cal B}(\pi^0 K^+)}{{\cal B}(\pi^- K^+)} \frac{\tau_{B^0}}{\tau_{B^+}} - 
    {\cal A}_{CP}(\pi^+ K^0) 
    \frac{{\cal B}(\pi^+ K^0)}{{\cal B}(\pi^- K^+)}  \frac{\tau_{B^0}}{\tau_{B^+}} - 
  } 
  & 
  \hspace{30mm}
  \\
  \hspace{30mm}
  &
  \multicolumn{2}{r}{
    {\cal A}_{CP}(\pi^- K^+) + 
    2 {\cal A}_{CP}(\pi^0 K^0) 
    \frac{{\cal B}(\pi^0 K^0)}{{\cal B}(\pi^- K^+)} = 0 \, .
  }
\end{array}
\en
As before, $\Delta$ represents the partial width difference,
${\cal A}_{CP}$ the direct $CP$ asymmetry and 
${\cal B}$ the ($b$ and $\bar{b}$ averaged) branching fraction.
This sum rule is derived using isospin conservation.
Since EWP do not respect isospin, their presence in the $K\pi$
modes was expected to cause a violation of the sum-rule. 
However, recently Gronau~\cite{mg05} has argued that 
possible violations due to SM EWP are extremely small,
making this a very precise test of the SM,
with uncertainty much below $1\%$.
For other relevant recent discussions, see~\cite{sum_rule}.
Note also that a similar sum rule for the decay rates exists~\cite{hl98,gr99}.

The current experimental situation,
including results presented at ICHEP 2006, is~\cite{hfag,kpiICHEP2006}
\begin{equation}
  \begin{array}{rcl@{\hspace{5mm}}rcl}
    {\cal A}_{CP}(\pi^0 K^+) & = & \ph{-}0.047 \pm 0.026 \, , & 
    {\cal B}(\pi^0 K^+) & = & ( 12.8 \pm 0.6 ) \times 10^{-6} \, , \\
    {\cal A}_{CP}(\pi^+ K^0) & = & \ph{-}0.009 \pm 0.025 \, , & 
    {\cal B}(\pi^+ K^0) & = & ( 23.1 \pm 1.0 ) \times 10^{-6} \, , \\
    {\cal A}_{CP}(\pi^- K^+) & = & -0.093 \pm 0.015 \, , & 
    {\cal B}(\pi^- K^+) & = & ( 19.7 \pm 0.6 ) \times 10^{-6} \, , \\
    {\cal A}_{CP}(\pi^0 K^0) & = & -0.12 \pm 0.11 \, , & 
    {\cal B}(\pi^0 K^0) & = & ( 10.0 \pm 0.6 ) \times 10^{-6} \, . \\
    \multicolumn{6}{c}{
      \frac{\tau_{B^+}}{\tau_{B^0}} =  1.076 \pm 0.008
    }
  \end{array}
\end{equation}
So, neglecting correlations,
\begin{equation}
  \begin{array}{l@{\hspace{8mm}}c@{\hspace{8mm}}r}
    \multicolumn{2}{l}{
      \Sigma\left({\cal A}_{CP}(\pi K)\right) = } \\
    & 
    \multicolumn{2}{r}{
      \left( 
        2 \times ( 3.3 \pm 1.8 ) - ( 1.1 \pm 3.1 ) - ( -9.3 \pm 1.5 ) + 2 \times ( -6 \pm 6 )
      \right) \times 10^{-2} = \left( 3 \pm 8 \right) \times 10^{-2} 
    }
  \end{array}
\end{equation}
  
The largest contribution to the uncertainty 
comes from the statistical error in the measurement of ${\cal A}_{CP}(\pi^0 K^0)$.
This is large because the reconstructed final state for this mode ($\pi^0 K_S$)
is a $CP$ eigenstate and contains no information about the flavour 
of the decaying $B$ meson.
This information must therefore be obtained using flavour tagging,
which in general can be performed with much higher efficiency at an
$e^+e^-$ $B$ factory than at a hadron collider.
Furthermore, this mode will be difficult to reconstruct 
in a hadronic environment, and time-dependent studies,
which are used by the $B$ factories to optimize the sensitivity,
are impossible since the vertex location cannot be reconstructed.

As mentioned, the above values neglect correlations between 
systematic effects in the various ${\cal A}_{CP}$ measurements.
This is a rather crude treatment, though not unreasonable while 
statistical errors dominate.  
In future, more precise results for the sum rule should be provided 
directly by the experiments, since systematic correlations can 
easily be taken into account in the analyses.

Within the Standard Model, the sum rule should hold to at least an 
order of magnitude below the current experimental error.
Precise tests of this relation can only be achieved with a Super $B$ Factory.

\subsection{\boldmath
  Direct $CP$ in $\pi^+ \pi^0$}

Since $\pi^+ \pi^0$ is an $I=2$ final state it cannot receive any contribution
from QCD penguins as they carry $\Delta I=0$ and therefore
can only contribute to an $I=1/2$ final state in $B^+$ decays.
Thus this final state can only receive contributions from trees and EWP. 
Consequently the SM can only give rise to extremely small 
direct $CP$ asymmetry in any such $I=2$ final state.
Explicit calculations show that while FSI cause
the asymmetry to increase substantially from
purely short distance expectations of around $5 \times 10^{-5}$, 
the resulting asymmetry $-0.009^{+0.002}_{-0.001}$~\cite{ccs1} 
stays below a few percent.
Moreover, there are arguments that any such asymmetry should be 
orders of magnitude smaller than this~\cite{gpy1999}. 
A significantly larger asymmetry would be a strong signal of NP.

The current average of the $B$ factory results is 
${\cal A}_{CP}(B^+ \to \pi^+ \pi^0) = 0.04 \pm 0.05$~\cite{hfag}.
Therefore, it is expected that the experimental precision 
of these results will approach the level of theoretical uncertainty 
at the end of the current generation of $B$ factory experiments.
This should motivate further theoretical studies of this channel.

\subsection{\boldmath 
  Time-dependent $CP$ asymmetries in exclusive radiative $B$ decays}

In 1997 an important test~\cite{ags} of the SM was
proposed which used (mixing-induced) time-dependent $CP$ violation (TDCP)
in exclusive modes such as $B^0 \to K^* \gamma$ and $B^0 \to \rho \gamma$.
This is based on the simple observation that, in the SM, 
photons produced in radiative $B^0$ decays are predominantly right-handed 
whereas those in $\bar{B}^0$ decays are predominantly left-handed. 
To the extent that the final states of $B^0$ and $\bar{B}^0$ decays
are different, TDCP would be suppressed in the SM.
Recall, the leading order effective Hamiltonian,
for $q = s,d$, can be written as
\be
H_{\rm eff} = - \sqrt{8} G_F \frac{e m_b}{16\pi^2} F_{\mu\nu}
\left [
  F_L^q\ \OL q\sigma^{\mu\nu}\frac{1+\gamma_5}{2}b
  +
  F_R^q\ \OL q\sigma^{\mu\nu}\frac{1-\gamma_5}{2}b
\right ] + h.c.
\label{effective_H}
\en

\noindent 
Here $F_L^q$ ($F_R^q$) corresponds to the amplitude for the
emission of left (right) handed photons in the $b_R \to q_L \gamma_L$
($b_L \to q_R \gamma_R$) decay, {\it i.e.} in the $\bar{B} \to \bar{F} \gamma_L$
($\bar{B} \to \bar{F} \gamma_R$) decay.
Based on the SM {\it leading order} $H_{\rm eff}$ 
in $b$ quark decay ({\it i.e.} $\bar{B}$ decays), 
the amplitude for producing wrong helicity (RH) photons 
$\propto F_R^q/F_L^q \propto m_q/m_b$,
where $m_q = m_s (m_d)$ for $b \to s\gamma (b \to d\gamma)$.
Consequently, time-dependent $CP$ asymmetries,
which occur when the final state is accessible to both $B$ and $\bar{B}$,
are suppressed by $\approx m_q/m_b$.
More detailed treatments, including effects of QCD corrections,
give the level of suppression 
as $\Lambda_{\rm QCD}/m_b$~\cite{gglp,gp2005}.
The most recent calculations show that the asymmetry
is indeed small in the Standard Model:
a perturbative QCD calculation gives $(-3.5 \pm 1.7)\%$~\cite{ms2006}
while a QCD factorization based approach gives 
$(-2.2 \pm 1.5 \,^{+0.0}_{-1.0})\%$~\cite{bz2006}.

Interestingly, emission of wrong-helicity photons from
$B$ decays is not suppressed in many extensions of the SM.
For example, in Left-Right Symmetric models (LRSM),
SUSY~\cite{chn03,okada03,chl00}
or Randall-Sundrum (warped extra dimension~\cite{aps04}) models,
in fact they can be enhanced by the ratio $m_{\rm heavy}/m_b$ where
$m_{\rm heavy}$ is the mass of the virtual fermion in the penguin-loop.
In LRSM as well as some other extensions this enhancement can
be around $m_t/m_b$. 
So while in the SM the asymmetries are expected to be very small, 
they can be sizeable in LRSM~\cite{ags}
as well as in many other models.

An important generalization was made in Ref.~\cite{aghs}. 
It was shown that the basic validity of this test of the SM does not require
the final state to consist of a spin one meson 
(a resonance such as $K^*$ or $\rho$) in addition to a photon.
In fact the hadronic final states can equally well be two mesons; 
{\it e.g.} $\ks (\pi^0, \eta^\prime, \eta, \phi...)$ or $\pi^+ \pi^-$.
Inclusion of these non-resonant final states, in addition to the
resonances clearly enhances the sensitivity of the test considerably.
For the case when the two mesons are antiparticles,
{\it e.g.} $\pi^+ \pi^-$ the angular distribution must be studied.
Although this analysis is more complicated, 
the outcome is that both the magnitude and the weak phase 
of any new physics contribution may be determined~\cite{aghs}.

In principle, photon emission from the initial light-quark
is a non-perturbative, long-distance, contamination
to the interesting signal of the short-distance
dipole emission from $H_{\rm eff}$~\cite{abs,grinpir}. 
Fortunately, it can be shown~\cite{aghs} that predominantly these LD
photons have the same helicity as those from $H_{\rm eff}$.
Another important source of SM contamination was recently emphasized in
Ref.~\cite{gglp} from processes such as $b \to s \gamma + $ gluon
which are from non-dipole operators. Such processes do not fix the
helicity of the photon and so can make a non-vanishing SM
contribution to mixing induced $CP$.

It was emphasized in Ref.~\cite{aghs} that the presence of such 
non-dipole contributions can be separated from the dipole contributions.
Although it requires a larger amount of data, 
the resolution to this problem is data driven.
To briefly recapitulate,
the different operator structure in $H_{\rm eff}$
would mean, that in contrast to the pure dipole case,
the time-dependent $CP$ asymmetry ($S$) would be a function of the
phase space (often a Dalitz plot) of the final state.
Thus, the contamination from non-dipole terms in $H_{\rm eff}$ may be
subtracted by fitting the experimental data on $S$ to
a suitable parametrization of the phase space dependence.
Note also that a difference in the values
of $S$ for two resonances of identical $J^{PC}$ 
could also indicate the presence of non-dipole contributions. 

To date, experimental results exist only for the decay
$B^0 \to \ks\pi^0\gamma$~\cite{belle-kspi0gamma,babar-kspi0gamma}.
The average is $S_{\ks\pi^0\gamma} = -0.28 \pm 0.26$~\cite{hfag}.
Similar to the hadronic penguin modes,
above $5 \times 10^{10}$ $B\bar{B}$ pairs are required to reduce
the uncertainty to the few percent level necessary for this ANT.
Larger data samples would be necessary for precise studies of the 
Dalitz plot dependence of the $CP$ asymmetry.
As is the case for $B^0 \to \ks\pi^0$, 
this measurement can only be achieved at an $e^+e^-$ $B$ factory,
where the $B$ decay vertex can be reconstructed using knowledge of the 
interaction region.
For other related radiative $b \to s$ decay channels, such as 
$B^0 \to \ks\eta^{(\prime)}\gamma$ and $B^0 \to \ks\phi\gamma$~\cite{otherkgamma},
the statistics are not yet sufficient to allow time-dependent analyses.
These modes would, however, be accessible at a Super $B$ Factory.
Similarly, analyses on radiative $b \to d$ decays,
such as the recently observed 
$B^0 \to \rho\gamma$ and $B^0 \to \omega\gamma$~\cite{belle-rhogamma,babar-rhogamma} 
would also become feasible \textendash{} these are likely to remain statistics 
limited even with very high luminosities.
Some of the above modes 
(those with tracks originating from the $B$ decay vertex)
can also be studied in a hadronic environment,
where the production of $B_s$ mesons will also facilitate the 
time-dependent study of $B_s \to \phi\gamma$.

\subsection{\boldmath
  Transverse $\tau$ polarization in semileptonic decays}

Precise measurements of rates and asymmetries in semileptonic $B$ decays 
can provide clean tests of the Standard Model.
In particular, decays with $\tau$ mesons are useful to probe the Higgs sector,
and have not yet been well studied.
A non-vanishing value of the $CP$-odd transverse polarization, $p^t_l$,
of charged leptons in semileptonic decays 
is a clean test of the SM~\cite{abes_pr,aes93,zl95}. 
This observable is odd under 
naive time reversal symmetry ($T_N$)~\cite{abes_pr},
and as such (due to the $CPT$ theorem) does not require a strong phase.
Thus it can arise from tree graphs in BSM scenarios 
{\it e.g.} if charged Higgs particles carry non-standard phase(s). 
Note that, in principle, non-vanishing $p^t_l$ can arise from 
$CP$-even final state interactions.
Therefore, for $<p^t_l> \neq 0$ to be a genuine $CP$-odd signal,
a comparison of positive and negative charged leptons is necessary.

The above argument applies to any semileptonic decay,
and, for example, is of interest to search for $CP$ violation in
top quark decays~\cite{abes_pr}.
Note also that searches for transverse muon polarization 
in kaon decays have been performed~\cite{kaon-pt}.
Any charged lepton ($e$, $\mu$ or $\tau$) can be used for such searches,
though here for practical reasons we focus on the $\tau$, 
whose decays serve as self-analyzers of its polarization.
Measurement of muon and electron polarization is impractical 
at any existing or planned $B$ facility.
Besides, in many BSM scenarios with an extended Higgs sector
the Higgs couplings to $\tau$ tend to be much bigger than
to the other charged leptons.
Specifically, one may use 
$[p^t_{\tau} \equiv S_{\tau} \dot p_{\tau} \times p_X] / [|p_{\tau} \times p_X|]$.  
Here, $S_{\tau}$ and $p_{\tau}$ are the spin and momenta of the $\tau$ 
and $p_X$ is the momentum of the final state hadron. 

The experimental detection of $p_{\tau}^t$ via decay correlations in $\tau$ decays
is highly challenging.
Only one similar analysis has yet been performed at the $B$ factories \textendash{}
a search for the $\tau$ electric dipole moment by Belle~\cite{belle-tauedm}.
Semitauonic $B$ decays result in at least two neutrinos in the final state,
and are therefore best studied using kinematic constraints from the 
reconstruction (in a hadronic final state) of the other $B$ in the event.
Previous studies~\cite{superb} have shown that rates and distributions for
$B \to D\tau\nu$ can be studied with ${\cal O}(10^{10}$ $B\bar{B}$) pairs 
produced at an $e^+e^-$ $B$ factory.
Clearly, energy or rate asymmetries in these semileptonic 
decays should also be studied.
However, one expects that in a charged Higgs scenario 
the transverse polarization asymmetry will be considerably bigger~\cite{bs04}.
Since the sensitivity to this observable is smaller,
higher luminosities will be required to make the necessary precise studies.
This then calls for a Super $B$ Factory.

As mentioned before, the study of transverse polarization in
semileptonic decays is also a very important test for use
in top decays (say): $t \to b \tau \nu$.
At the LHC this should be a very sensitive test for searching for
$CP$-odd phase from a charged Higgs exchange~\cite{aes93,abes_pr}.   
Although the study of semileptonic $B$ decay at the LHC looks challenging,
these processes need to be carefully studied.

\section{Summary}

\begin{table}
  \begin{tabular}{|c|c|c|c|}
    \hline
    Observable & SM expectation & 
    Current expt. uncertainty & 
    $B\bar{B}$ pairs needed \\  
    & & ($B\bar{B}$ pairs used) & \\
    \hline
    $\Delta S[\eta^\prime K^0,\phi K^0,K^0\bar{K}^0K^0,\ldots]$ & 
    $\sim (0 \pm 2)\%$ & $20\%$ ($6 \times 10^8$)& $5 \times 10^{10}$ \\
    \hline
    ${\cal A}_{CP}^{s+d}[M^0X_{s+d}]$ & 
    $\lsim 0.1\%$ & \textemdash{} &$> 10^{12}$  \\
    \hline
    ${\cal A}_{CP}[X_s\gamma]$ & 
    $(0.5 \pm 0.2)\%$ & $4\%$ ($2.4 \times 10^8$) & $10^{11}$ \\
    ${\cal A}_{CP}[X_d\gamma]$ & 
    $(-10 \pm 5)\%$   & \textemdash{} & $10^{11}$ \\
    ${\cal A}_{CP}[X_{s+d}\gamma]$ & 
    $(0.000 \pm 0.001)\%$ & $12\%$ ($10^8$) & $> 10^{12}$ \\
    \hline
    ${\cal A}_{CP}[X_s l^+ l^-] $ & 
    $(-0.2 \pm 0.2)\%$ & $26\%$ ($10^8$) & $10^{12}$   \\
    ${\cal A}_{CP}[X_d l^+ l^-] $ &
    $(4 \pm 4)\%$  & \textemdash{} & $10^{12}$   \\
    ${\cal A}_{CP}[X_{(s,d)}l^+ l^-] $ &  & \textemdash{} & $> 10^{12}$   \\
    \hline
    $\Sigma\left({\cal A}_{CP}(\pi K)\right)$ & 
    $(0 \pm 1)\%$ & $15\%$ ($6 \times 10^8$) & $> 10^{11}$ \\ 
    \hline
    $A_{CP}(\pi^+ \pi^0)$ & 
    $\lsim 1\% $ & $6\%$ ($6 \times 10^8$) & $10^{10}$\\    
    \hline
    $S[K_S \pi^0 \gamma, \ldots ]$ & 
    $\sim (0 \pm 5)\% $ & $28\%$ ($6 \times 10^8$) & $> 10^{10}$  \\
    \hline
    $<p_t^{\tau}>(D(X_c) \tau \nu_{\tau})$ & 0 & \textemdash{} & $> 10^{12}$ \\
    \hline
  \end{tabular}
  \caption{ 
    Illustrative sample of {\it approximate null tests (ANTs)}, 
    with rough SM expectations and theory errors,
    current experimental uncertainties and 
    estimates of numbers of $B$ mesons needed 
    for a Super $B$ Factory
    to approach the SM uncertainty.
    More details for each mode can be found in the text.
  } 
\label{ants}
\end{table}

The $B$ factories have performed superbly and have improved our
understanding of $CP$ violation phenomena remarkably.
The CKM paradigm of $CP$ violation has been confirmed as 
the dominant source of the observed $CP$ violation.
Beyond the SM $CP$-odd phase(s) are likely to cause small
perturbation in $B$ physics. 
Null tests of the SM may be very useful to search for small effects; 
indeed, the stricter the null tests the better it is.  

We have presented a brief discussion of several important null tests. 
The effective search for many of these (see Table~\ref{ants}
for an illustrative sample of the many clean ANTs) will require 
the very large samples of clean $B$ mesons that can only 
be produced at an $e^+e^-$ based Super $B$ Factory.
In fact, some of these ANTs will remain statistics limited even
with Super $B$ Factory statistics.  
The key point is that by pushing as close as possible
to the SM theoretical error, 
these measurements will provide effective searches for NP effects.
In this paper we have examined only a fraction 
of the ANTs which may provide sensitive tests of the Standard Model.
In addition to those discussed, there is also the possibility 
to search for new flavour changing neutral currents in both the 
lepton ({\it e.g.} $\tau \to \mu\gamma$, $\tau \to \mu\mu\mu$, {\it etc.}) and
quark ({\it e.g.} $b \to ss\bar{d}$, $b \to dd\bar{s}$~\cite{fkk2006}) sectors.
Once NP has been discovered, diagnosis of its origin will require the 
correlated study of many such observables, 
as well as those from other experiments~\cite{superb}.

While we have concentrated on the ANTs that are best examined at a 
Super $B$ Factory, there are several excellent tests of the 
Standard Model that will be put under the 
microscope at the LHCb experiment in coming years.
These include one particularly noteworthy ANT:
the search for sizeable $CP$ violation in the 
$B^0_s-\bar{B}^0_s$ mixing phase $\phi_s$,
measured using $B^0_s \to J/\psi \phi$.
The results from LHCb will be extremely important to test the Standard Model.
However, a Super $B$ Factory is necessary to fully exploit the 
sensitivity to new physics of flavour parameters. 

A global effort to build the optimum Super $B$ Factory machine therefore
appears very timely and worthwhile. 
At present, different accelerator schemes are under consideration~\cite{isbf}.
Here, we will not discuss the relative merits of these.
Either machine would be suitable for the purpose of 
studying the channels discussed in this paper,
so long as the luminosity is high enough, 
the environment is clean enough, 
and the experiment takes place soon enough!

The International Super $B$ Factory will be highly relevant in the LHC era.
It will enable tests of the Standard Model that are 
wholly complementary to those performed at the energy frontier.
Indeed, the many precise measurements which can be carried out
at the International Super $B$ Factory therefore
should significantly extend the reach of the LHC.

\vskip 2.0cm \acknowledgments  
We are grateful to Michael Gronau and Jure Zupan for very useful discussions.
This research was supported in part by 
the U.S. DOE contract No. DE-AC02-98CH10886(BNL).

\end{document}